\documentclass[twocolumn,prd,nofootinbib,aps,prl,floats,floatfix,superscriptaddress,amsmath,amssymb,longbibliography,secnumarab
ic]{revtex4-1} %

\usepackage[final]{graphicx}
\usepackage{hyperref}
\usepackage{amsmath}
\usepackage{bbm}
\usepackage{amsfonts}
\usepackage{amssymb}
\usepackage{latexsym}
\usepackage{graphicx}
\usepackage[english]{babel}
\usepackage{multirow}
\usepackage{float}
\usepackage{url}
\usepackage{slashed}
\usepackage{xcolor} 
\usepackage[utf8]{inputenc}
\usepackage{verbatim}
\usepackage{stackengine}
\usepackage{cancel}
\usepackage{accents}
\usepackage{array}
\usepackage{physics}
\usepackage{bm}

\def\sfrac#1#2{{\textstyle{#1\over #2}}}

\newcommand{\be}{\begin{equation}}
\newcommand{\ee}{\end{equation}}
\newcommand{\ba}{\begin{array}}
\newcommand{\ea}{\end{array}}
\newcommand{\bea}{\begin{eqnarray}}
\newcommand{\eea}{\end{eqnarray}}

\usepackage{slashed}

\begin{document}
\rightline{CERN-TH-2024-154}

\title{Plausible constraints and inflationary production for dark photons}
\author{James M.\ Cline}
\email{jcline@physics.mcgill.ca}
\thanks{ORCID: \href{https://orcid.org/0000-0001-7437-4193}{0000-0001-7437-4193}}
\affiliation{McGill University Department of Physics \& Trottier Space Institute, 3600 Rue University, Montr\'eal, QC, H3A 2T8, Canada}
\affiliation{CERN, Theoretical Physics Department, Geneva, Switzerland}
\author{Gonzalo Herrera}
\email{gonzaloherrera@vt.edu}
\thanks{ORCID: \href{https://orcid.org/0000-0001-9250-8597}{0000-0001-9250-8597}}
\affiliation{Center for Neutrino Physics, Department of Physics, Virginia Tech, Blacksburg, VA 24061, USA}
\affiliation{CERN, Theoretical Physics Department, Geneva, Switzerland}

\begin{abstract}
Generic constraints on dark photons are generally presented assuming they have Stueckelberg masses.  These constraints are strengthened if instead the mass is due to the Higgs mechanism and the dark Higgs is light.  First, we show that under reasonable assumptions on the origin of kinetic mixing $\epsilon$ and perturbativity, the strengthened constraints on $\epsilon$ cannot be arbitrarily relaxed by making the Higgs heavy.  Second,  we demonstrate a simple mechanism for generating dark photon dark matter after inflation, 
where fluctuations of a dark Higgs by stochastic misalignment can produce stable dark photons through $h\to A'A'$ decay. 
Third, we point out new  {\it lower} bounds on $\epsilon$ in the case where the dark photon mediates thermal freeze-out of light dark matter by $s$-channel
exchange, taking account of generic expectations for the size of $\epsilon$, and astrophysical upper bounds on the self-interaction cross section.  
\end{abstract}
\maketitle

\section{Introduction}
Dark photons $A'$ are amongst the simplest models of hidden sectors \cite{Holdom:1985ag},
and can be the dark matter itself, or a mediator explaining the relic density of another dark matter particle. $A'$ is expected to kinetically mix with the standard model photon with a 
strength $\epsilon \sim g e/(16 \pi^2)$, where $g$ is the hidden U(1) gauge coupling and $e$ is the electromagnetic charge.  Such 
mixings are mediated by loops of a heavy particle that carries both kinds of charge.   It is possible to avoid this level of mixing in models where different loop contributions cancel each other \cite{Obied:2021zjc,Hebecker:2023qwl}.
In this work we will focus on the generic situation where no such cancellation occurs.

Many laboratory and astrophysical bounds on $\epsilon$ versus $m_{A'}$ have been derived in the last decades,
which commonly assume that $A'$ gets its mass from the Stueckelberg mechanism.  For practical purposes, this is just introducing a bare mass term.  It is well-known that more severe constraints on $\epsilon$ can be derived if instead $A'$ gets its mass from a dark Higgs mechanism \cite{Ahlers:2008qc,An:2013yua}.  If the dark Higgs $h$ is sufficiently light, it can be emitted by stars and lead to anomalous cooling.  This is model-dependent, since one can evade such bounds by making $h$ sufficiently heavy, or the gauge coupling $g$ sufficiently small.  Here we make a simple but powerful observation: perturbative unitarity, combined with the previous assumption about kinetic mixing, puts an upper limit on the relevant mass ratio $m_{h}/m_{A'}$.  This allows for new model-independent constraints on $\epsilon$ versus $m_{A'}$ in the case of the dark Higgs mechanism, which we derive in Section \ref{higgsed-sect}.

If light $A'$ constitutes the dark matter, it is difficult to explain its relic density in terms of the conventional misalignment mechanism,
since the conformal properties of vectors cause such perturbations to damp more quickly with the Hubble expansion than cold dark matter.  It has been proposed that $A'$ could inherit its relic density by being the decay product of a (pseudo)scalar particle 
such as an axion, 
whose inflationary fluctuations do not suffer such damping
\cite{Agrawal:2018vin,Co:2018lka}.  

\begin{figure*}[t]
\centerline{
\includegraphics[scale=0.6]{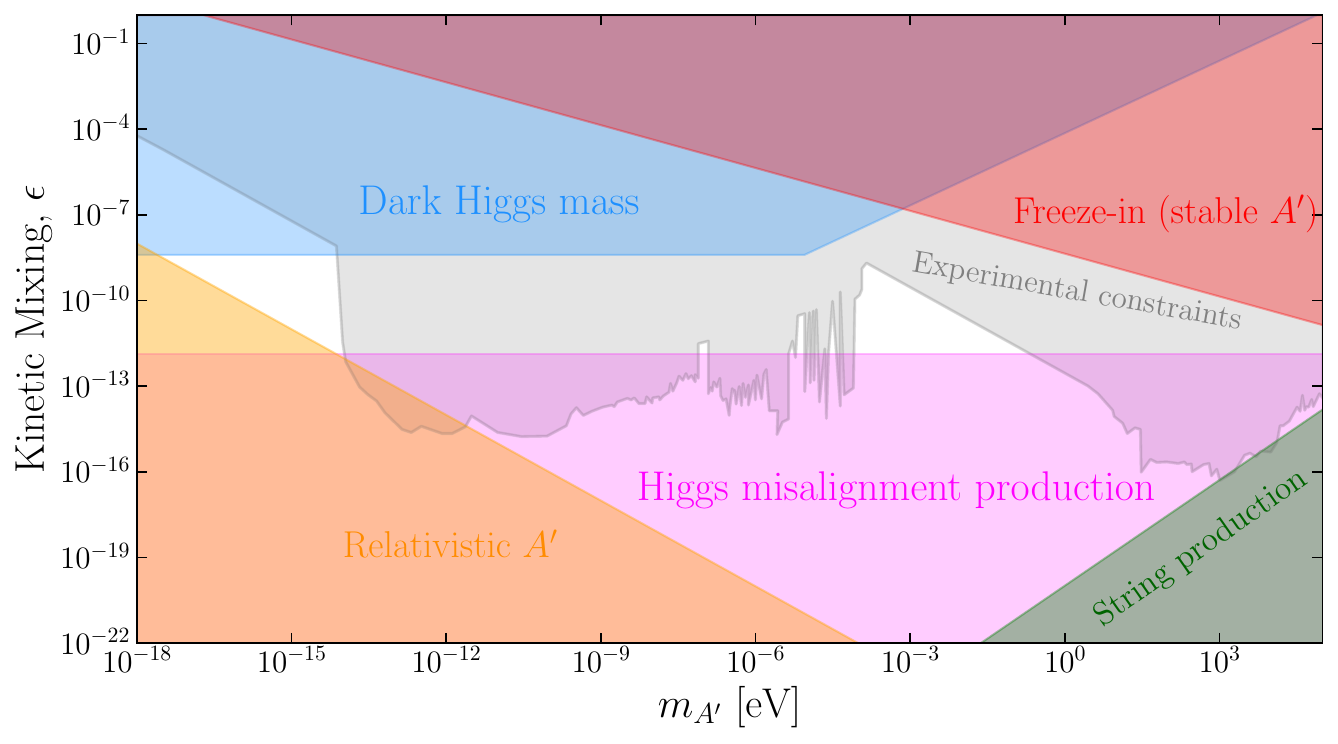}
}
\vspace{-0.2cm}
\caption{Cyan: excluded region from stellar cooling plus perturbative unitarity of dark Higgs self-coupling, derived in Section \ref{higgsed-sect}.  
black: region excluded by overproduction of $A'$ {from the decays of inflationary misaligned dark Higgs}, derived in Section \ref{sect:infl}. Red: Excluded region due to overproduction of $A^{\prime}$ via freeze-in, derived in Section \ref{sect:infl}. \textcolor{black}{Orange: Region of parameter space where the decays $h \rightarrow A^{\prime}A^{\prime}$ yield relativistic dark photons today.}  Grey: Combination of astrophysical, cosmological and laboratory constraints on dark photons, adapted from Ref.\ \cite{Caputo:2021eaa}. {Green: Region excluded by cosmic string formation, from Ref.\ \cite{Long:2019lwl}}.}
\label{fig:dp}
\end{figure*}

In the present context, a more economical model is to instead use the dark Higgs since it is already present to generate the $A'$ mass. \textcolor{black}{After shifting the dark Higgs $h$ by its VEV, 
$h\to h+v$, and using $v = m_{A'}/g$ to eliminate $v$, the Lagrangian is
\begin{align}
\mathcal{L} = & -\frac{1}{4}F_{\mu \nu}F^{\mu \nu} - \frac{1}{4}F^{\prime}_{\mu \nu}{F^{\prime}}^{\mu \nu} - \frac{\epsilon}{2}F_{\mu \nu}{F^{\prime}}^{\mu \nu} \notag \\
& + \frac{m_{A^{\prime}}^2}{2}A^{\prime}_{\mu}{A^{\prime}}^{\mu} + {A^{\prime}}^{\mu} g_{\chi} \bar{\chi} \gamma_\mu \chi \notag \\
& + g m_{A^{\prime}} h A^{\prime}_{\mu}{A^{\prime}}^{\mu} + \frac{1}{2}g^2 h^2 A^{\prime}_{\mu}{A^{\prime}}^{\mu}.
\end{align}
where ${F^{\prime}}^{\mu \nu}=\partial^\mu {A^{\prime}}^{\nu}-\partial^\nu {A^{\prime}}^{\mu}$, and $\chi$ is a dark fermion. The term involving $\chi$ will only become relevant in Section \ref{sect:relic}, where the dark photon will be considered as a mediator.}  
Ref.\ \cite{Dror:2018pdh} showed that if $h$ is displaced from the minimum of its potential at the end of inflation, then $A'$ can obtain the desired relic density by parametric resonant production.  That study assumed the initial value of $h$ could be arbitrary.  Here we note that stochastic growth of fluctuations during inflation gives rise to a calculable estimate for this starting value, in terms of the inflationary energy scale
\cite{Chatrchyan:2022dpy}.  Moreover, simple perturbative decays $h\to A'A'$ are sufficient to give the desired relic density.  Overproduction of $A'$ leads to new constraints on $\epsilon$; this is discussed in Section \ref{sect:infl}.

Lastly, we consider $A'$ in the role of mediator between dark matter $\chi$ and standard model fermions, via kinetic mixing. This scenario has been widely considered in the literature as a thermal benchmark model for sub-GeV dark matter \cite{Boehm:2003hm,Essig:2011nj, Essig:2022dfa}.   We point out that in the generic situation where $m_{A'}>m_\chi$, a lower bound on
$\epsilon$ arises from overclosure of the Universe.  This excludes an interesting region of parameter space if one adopts our working assumption $\epsilon\sim g_\chi e/(16\pi^2)$.  Additionally taking account of Bullet Cluster constraints on the dark matter self-interactions, we further constrain these regions, as discussed in 
Section \ref{sect:relic}.

\section{Higgsed Dark photon constraints}
\label{higgsed-sect}

If a dark Higgs $h$ gives rise to the dark photon mass by getting a VEV $v$, it can participate in processes that would otherwise only involve $A'$ \cite{Ahlers:2008qc}.   The effect is important when $m_{A'}$ is negligible compared to the energy scales $T$ of interest.  Then one can work in the basis where $h$ acquires a millicharge and can thereby be emitted from stars, leading to anomalous cooling by electromagnetic processes, \textcolor{black}{mainly driven by the plasmon decay channel $\gamma^{\star} \rightarrow A^{\prime}+h$}.  To avoid ensuing bounds  \cite{Davidson:2000hf}, one must assume that $m_h > T$ so that the emission is kinematically blocked.  

However, there is a theoretical upper limit on $m_h^2 = 2\lambda v^2$, when written
in terms of the VEV and self-coupling $\lambda$ in the potential
$V = \lambda(|H|^2-v^2/2)^2$, where $H$ represents the complex field.    Perturbative unitarity of $\lambda$ implies that $\lambda < 8\pi/3$, while $v$ is related to the dark photon mass by $m_{A'} = g v$. These can be tied together by making a generic assumption about the size of the kinetic mixing, in the absence of any special cancellations: \textcolor{black}{the typical contribution from a loop of a heavy particle $\phi$ carrying both U(1) charges gives \cite{Pich:1998xt}
\be
\epsilon=\frac{g e}{16 \pi^2} \mathrm{log}\Big (\frac{m_{\phi}^2}{\mu_{0}^2} \Big )
\ee
where $\mu_{0}$ is the renormalization scale. } 

The logarithm could accidentally be small, or there could be a cancellation mechanism,
such as two nearly equal-mass particles with opposite charges 
contributing in the loop.  In such cases we do not gain any new constraints from the consideration of kinetic mixing.  In the present work we are interested in the generic situation in which the
log is of order unity, and we henceforth set it to 1.

Then we can relate the kinetic mixing $\epsilon$ with the gauge coupling $g$
\be \label{eq:size_epsilon}
    \epsilon \sim {g e\over 16\pi^2}\,.
\ee
Further eliminating $v,\,g,\,m_h$ in favor of $m_{A'},\,\epsilon,\,T$, we obtain
\be
   \epsilon < \sqrt{1\over 3} {e\, m_{A'}\over 4\pi^{3/2}\, T} 
\ee
in order to avoid $h$ emission by stars of temperature $T\sim 1$\,keV.
\textcolor{black}{This bound applies in the region of parameter space where it can soften a previously derived limit for sufficiently light $h$ \cite{An:2013yua}. That reference obtained
constraints of order $\epsilon \lesssim 10^{-13}$ from the anomalous cooling of horizontal branch stars,
but this relied on the {\it ad hoc} assumption that $g=0.1$. Refined stellar cooling constraints from the tip of the red-giant branch lie on the same order of magnitude \cite{Fung:2023euv}. When relating $g$ and $\epsilon$ via Eq. \ref{eq:size_epsilon}, we find instead the limit
\begin{equation}
\epsilon \lesssim \sqrt{\frac{8 \times 10^{-15}e}{16 \pi^{2}}} \lesssim 3.9 \times 10^{-9}.
\end{equation}
The scaling of the bound in $\epsilon$ with $g$ arises from the plasmon decay rate governing the energy loss in stars, which scales as $\epsilon^2 g^2$ \cite{Raffelt:1996wa}.  Our combined stellar cooling constraints are shown in Fig.\ \ref{fig:dp} as the cyan region.}

\section{Inflationary generation of $A'$ via dark Higgs}
\label{sect:infl}

Ref.\ \cite{Dror:2018pdh} showed how parametric resonance of an oscillating dark Higgs $h$ could efficiently produce $A'$ dark matter.  Their analysis relied upon arbitrary 
starting values of $h$ away from its minimum.  We note that a definite
initial value can be predicted from the
 stochastic misalignment of $h$ that gets generated by its fluctuations during inflation.
Furthermore we employ the simpler mechanism of perturbative decays
$h\to A' A'$ to produce the dark matter.  We find that this works in the regime where kinetic mixing is sufficiently small so that $A'$ is long-lived compared to the age of the Universe.  On the other hand, $h\to A' A'$ decays are unsuppressed 
for small $\epsilon$ or $g$; in the limit $g\to 0$, the width goes as 
$\Gamma \simeq g^2 m_h^3/(8 \pi m_{A'}^2) = \lambda m_h/(4\pi)$ due to the enhanced coupling to longitudinal $A'$s \cite{Araki:2020wkq}. \textcolor{black}{For the parameters of interest, this leads to decays well before big bang nucleosynthesis, except in the orange shaded region in Fig.\ \ref{fig:dp}, which is mostly excluded by other constraints. We note that a small triangular region remains viable, around $m_{A^{\prime}} \sim 10^{-17}-10^{-14}$ eV and $\epsilon \sim 10^{-12}-10^{-8}$}.

During inflation, $h$ acquires a nonzero value through its  fluctuations, with amplitude $\langle h^2\rangle \sim 3 H_I^4/(8 \pi^2 m_h^2)$ \cite{Graham:2018jyp,Chatrchyan:2022dpy},
where $H_I$ is the Hubble rate during inflation, $H_I^2 = \Lambda_I^4/(3 M_p^2)$, with $\Lambda_I$ being the energy scale.   This corresponds to an  energy density
\be
    \rho_i  = m_h^2 \langle h^2\rangle \sim {3 H_I^4\over 8\pi^2}\,.
\ee
This is only a mean value, and in some Hubble patches, $\langle h^2\rangle$ could take a smaller value, in analogy to the misalignment angle for axions accidentally being $\theta \ll 1$.  We leave aside this possibility here and simply assume the mean value.  
 If reheating is efficient, the entropy is initially $s_i\sim g_*\Lambda_I^3$.  We  estimate the abundance of $h$ as
\be
    Y_h \sim {\rho_i\over s_i m_h}\,.
\ee
Even if $h$ is strongly coupled to itself via $\sfrac14\lambda h^4$ with $\lambda \sim 8\pi/3$, this abundance should be roughly conserved. 

The simplest assumption is that the dark Higgs $h$ decays slowly through perturbative decays.  We can consider decays into $A'A'$ or into DM pairs
$\chi\bar\chi$ with no qualitative difference.  Suppose $A'$ is the DM,
as the simplest hidden sector.  We will verify that it is coupled so weakly that it does not thermalize.
Eventually $h$ decays to $2 A'$ and $A'$ inherits twice the $h$ abundance (neglecting $4\to 2$ annihilation processes).  Then to avoid overclosure by $A'$, we need
\be
    m_A' Y_{A'} \lesssim 4\times 10^{-10}\, {\rm GeV}\,.
\ee
Taking $g_*\sim 100$, this gives an upper limit on the scale of inflation,
\be
    \Lambda_I \lesssim 1.6\times 10^{14}\,{\rm GeV}  \left(m_h\over m_{A'}\right)^{1/5} \,.
    \label{Lambda-limit1}
\ee
Saturating this limit means that $A'$ constitutes the observed relic density.  One can alternatively use it to bound the scale of inflation.

\begin{figure*}[t]
    \centerline{
    \includegraphics[scale=0.6]{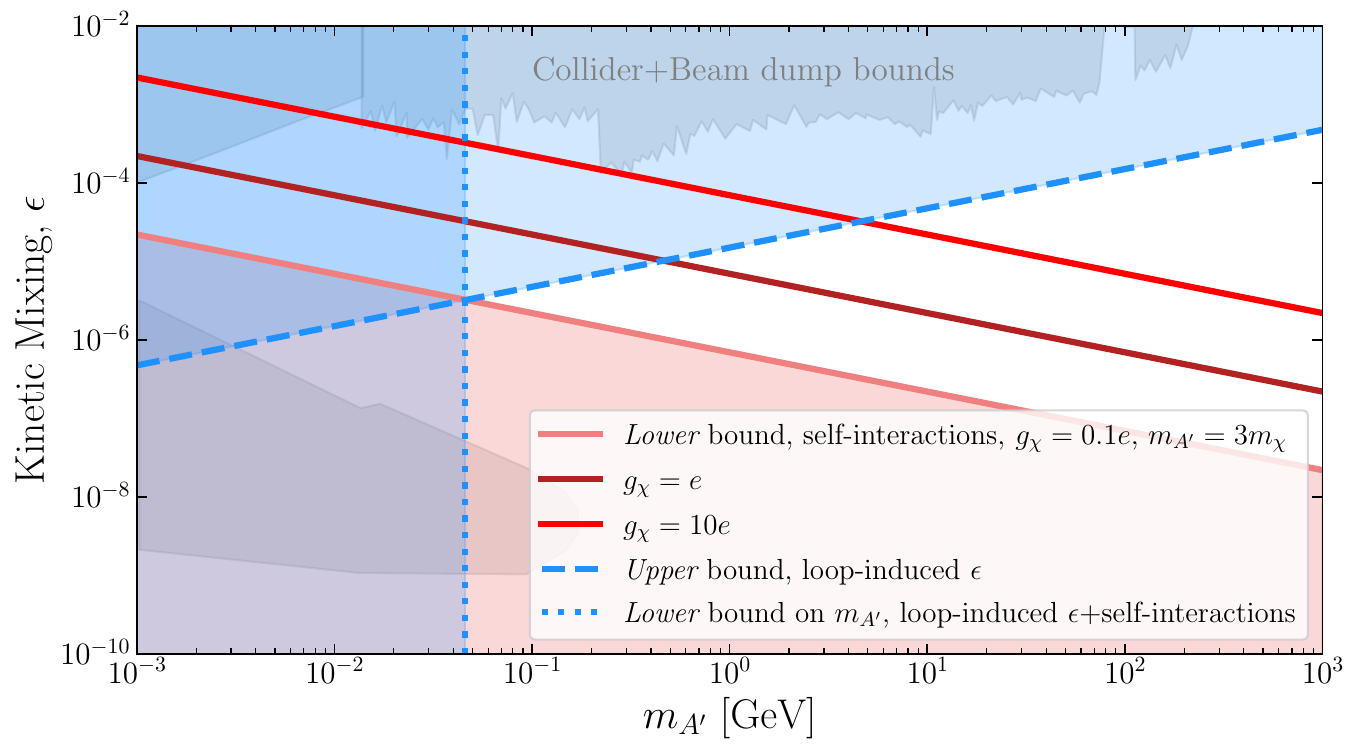}}
    \caption{Dashed line: Upper bound when $m_A' > m_\chi$, so that $\chi\chi\to A'\to f\bar f$ is the dominant annihilation process for thermal freezeout.
    Solid lines: Lower bounds on $\epsilon$ from $\chi\chi\to f\bar f$ dark matter freezeout and Bullet Cluster bound on $\chi\bar\chi\to\chi\bar\chi$ scattering, for assumed values of $g_\chi$ and $m_{\chi}/m_{A'} = 1/3$. Dotted line: Lower bound on the dark photon mass mediating thermal dark matter, from combining the self-interaction bound and the loop-induced estimate on the kinetic mixing}
    \label{fig:limit_epsilon_vs_mA}
\end{figure*}

In the spirit of section \ref{higgsed-sect}, we are motivated to 
consider strong higgs self-couplings $\lambda = 8\pi/3$, for weakening the limits on kinetic mixing, combined with the assumption $\epsilon \sim g e/(16\pi^2)$.  This gives
{\be
    {m_h\over m_{A'}}\lesssim\sqrt{\pi\over 3}{e\over 2\pi^2\epsilon} \simeq {0.014\over\epsilon} \,.
    \label{mass-ratio}
\ee}
On the other hand, Planck gets an upper limit on the inflation scale, $\Lambda_I< 1.6\times 10^{16}\,{\rm GeV}$ \cite{Planck:2018jri}.
Hence we cannot saturate the constraint (\ref{Lambda-limit1}) for $\epsilon \lesssim 10^{-12}$, in particular in the flat region of the allowed parameters from Section \ref{higgsed-sect}.  
To get the desired $A'$ relic density while satisfying (\ref{mass-ratio}), we need 
{\be
    \Lambda_I \lesssim 6.8\times 10^{13} \,
    \epsilon^{-1/5}\,{\rm GeV} \le 1.6\times 10^{16}\,{\rm GeV}\,.
    \label{Lambda-limit}
\ee}
This leads to the constraint {$\epsilon \gtrsim 1.4 \times 10^{-12}$}, which is
shown in Fig.\ \ref{fig:limit_epsilon_vs_mA} (black region).

We note that inflationary production of dark photons via stochastic misalignment motivates a region of parameter space currently inaccessible by direct detection experiments, cosmological and astrophysical probes, in the dark photon mass window from $10^{-6} \mathrm{eV} \lesssim m_{A^{\prime}} \lesssim 1 \mathrm{eV}$.
Our new mechanism is more efficient than freeze-in \cite{Hall:2009bx}, which
is not viable in the experimentally allowed region of parameter space.  We have checked this by estimating the cross section for production from 
photon scattering on electrons, $A e\to A' e$, as
$\langle\sigma v\rangle \sim \alpha^2\epsilon^2/(m_e^2 + T^2)$ at temperature $T$.  Integrating the Boltzmann equation for the $A'$
abundance \cite{Cline:2018fuq}
\be
    {dY_{A'}\over dx} ={x s \langle\sigma v\rangle \over H(m_{A'})}Y_{\rm eq}^2\,,
\ee
where $x = m_{A'}/T$, $s \sim g_* T^3$ is the entropy density, $H(m_{A'})$ is the Hubble rate at $T=m_{A'}$, and $Y_{\rm eq}\sim 1/g_*\sim 0.01$ is the equilibrium abundance, we find $Y_{A'}\sim (\pi/2) g_*^{-3/2} \alpha^2\epsilon^2 M_p/m_e$, and a resulting overclosure bound $\epsilon \gtrsim 10^{-8} ({\rm eV}/ m_{A'})^{1/2}$,
which lies in the excluded (grey) region of Fig.\ \ref{fig:limit_epsilon_vs_mA}.

To ensure $A'$ is metastable on the timescale of the age of the Universe, we assume that $m_{A'} < 2 m_e$, so the dominant decay channel is
$A' \to 3\gamma$, with rate $\Gamma \cong (\epsilon/0.003)^2(m_{A'}/m_e)^9$\,s$^{-1}$ \cite{Pospelov:2008jk,McDermott:2017qcg}.  This puts a very weak upper bound on $\epsilon$, which is satisfied everywhere in the allowed region of Fig.\ \ref{fig:limit_epsilon_vs_mA}.

A potentially efficient alternative mechanism for producing
dark photons is through the decay of a network of strings that form during the first-order phase transition when the dark U(1) symmetry breaks spontaneously
\cite{Redi:2022zkt}.  For small gauge coupling $g$, the symmetry is approximately global, and decay of $h$ into longitudinal $A'$s, or equivalently the Goldstone bosons, is unsuppressed by $g$.
\cite{Long:2019lwl}. This mechanism is efficient in the limit of large string tensions $\mu \cong v^2\ln^2(m_h/m_{A'})$, such that 
\be
    X \equiv \left(m_{A'}\over 10^{-13}\,{\rm eV}\right)^{1/2} {\mu\over (10^{14}\,{\rm GeV})^2} \gtrsim 1\,.
\ee
This is shown as the shaded lower right corner in Fig.\ \ref{fig:dp}.
For small couplings $\lambda < 10^{-3}$ , Ref.\ \cite{Redi:2022zkt} showed that
isocurvature fluctuations in the produced $A'$ particles can lead to strong constraints, but this is opposite to the regime of large $\lambda$ we focus on in the present work.

\section{Freeze-out by $s$-channel $A'$ exchange}
\label{sect:relic}

We next consider an enlarged dark sector in which a Dirac fermion $\chi$ is the dark matter, and the kinetically mixed dark photon is the mediator to the standard model, coupled to $\chi$ with strength $g_{\chi}$.  A lower bound on $\epsilon$
arises in the generic situation where  $m_{A'} > m_{\chi}$, so that
$\chi\bar\chi\to A'A'$ annihilations are kinematically blocked.
Then the annihilation process $\chi \bar{\chi} \rightarrow A' \rightarrow f\bar{f}$ sets the dark matter relic abundance. The velocity averaged annihilation cross section is given by
\begin{equation}
\langle\sigma v\rangle_{\rm ann} \simeq \frac{g_\chi^2 e^2\epsilon^2 m_\chi^2}{\pi m_{A'}^4}\,.
\label{eq:sigma_annihilation}
\end{equation}
In order to get the right relic density from thermal freezeout, 
we require \cite{Steigman:2012nb}
\begin{equation}
\langle\sigma v\rangle_{\mathrm{ann}} \simeq 4.4 \times 10^{-26} \mathrm{~cm}^3 / \mathrm{s} \simeq 1.5 \times 10^{-36} \mathrm{cm}^{2}\,,
\label{sig-ann-val}
\end{equation}
in units of $c=1$.

Assuming that $\epsilon\sim g_\chi e/16\pi^2$ as before, we can eliminate $g_\chi$ and obtain the upper bound
\be
    \epsilon \lesssim 2.6\times 10^{-5} \left(m_{A'}\over {\rm MeV}\right)^{1/2}
\ee
by combining Eqs.\ (\ref{eq:sigma_annihilation}-\ref{sig-ann-val}) with the working assumption
$m_{A'}> m_\chi$.  Since thermal freezeout requires $m_\chi\gtrsim 1\,$MeV, it only applies for heavier $A'$ than depicted in 
Fig.\ \ref{fig:dp}.  
In fact, we will show that $m_{A'}$ should exceed $\sim 50$\,MeV.
The excluded region is shown in Fig.\ \ref{fig:limit_epsilon_vs_mA}. 

In addition to annihilation, 
$t$-channel exchange of $A'$ unavoidably mediates elastic scattering between dark matter particles, with a cross section 
\begin{equation}
\sigma_{\rm el} \simeq \frac{g_\chi^4 m_\chi^2}{\pi m_{A'}^4}\,,
\end{equation}
that is unsupressed by $\epsilon$, in contrast to $\sigma_{\rm ann}$. Dark matter self-interactions in galaxy clusters constrain this scattering cross section at the level 
of \cite{Randall:2008ppe,Markevitch:2003at}
\begin{equation}
{\sigma_{\text {el}} \over m_{\mathrm{\chi}}} \lesssim 10^{-24}\, \mathrm{cm}^2 / \mathrm{GeV}\,,
\label{eq:sigma_scattering}
\end{equation}
which is much larger than the typical annihilation cross section.
Nevertheless, since one generically expects tat $\epsilon \ll g_{\chi}$, it is interesting to determine the minimum ratio $\epsilon/g_\chi$  compatible with the combined relic abundance and self-interaction constraints. Combining Eqs.\  (\ref{eq:sigma_annihilation}-\ref{eq:sigma_scattering}), we find that $\epsilon/g_\chi$ for thermally produced dark matter is bounded from below by
\begin{equation}
{\epsilon\over g_\chi} \gtrsim 4 \times 10^{-6}\left({\rm GeV}\over m_{\chi}\right)^{1/2}\,.
\end{equation}
One can satisfy this bound by taking $g_\chi$ to be sufficiently small.\footnote{The weak gravity conjecture implies that $g_\chi \gtrsim m_\chi/M_p$, but this gives a very weak constraint.}\ \ However if we reasonably estimate $\epsilon\sim e g/16\pi^2$, it gives constraint on the DM mass,
$m_\chi > 46\,$MeV, and by assumption, $m_{A'} > 46\,$MeV as well.

In the preceding, we assumed that $g_\chi = g = 16\pi^2 \epsilon/e$.
Conceivably $g_\chi$ could be much larger than $g$.  Considering representative values $g_\chi = 0.1\,e,\,e,\,10\,e$, and mass ratio $m_\chi/m_{A'} = 1/3$, we find interesting lower bounds on $\epsilon$ versus $m_{A'}$ illustrated in Fig.\ \ref{fig:limit_epsilon_vs_mA} (right).  This probes a region of parameter space which has not yet been excluded by collider or beam-dump experiments.

\section{Conclusions}
\label{sec:conclusion}

Dark photons ($A'$) have been widely studied, 
both as a well motivated dark matter candidate, and as a portal between the Standard Model and dark sectors. A variety of cosmological, astrophysical and laboratory probes restrict both of these scenarios by setting limits on the kinetic mixing $\epsilon$ between the dark photon and the Standard Model photon, but typically these invoke the simplifying assumption that 
dark photons get their mass via the Stueckelberg mechanism. In that scenario, large portions of the $\epsilon$ versus $m_{A'}$ plane remain open, whereas the constraints can be much more restrictive if a light Higgs $h$ in the dark sector generates the $A'$ mass.

In this paper we have derived the unavoidable stellar cooling constraints from
$h$ emission, using perturbative unitarity of the $h$ self-coupling to limit its mass,
and the reasonable expectation that $\epsilon$ is smaller than the
dark gauge coupling only by a loop suppression factor, $e/(16\pi^2)$,
to relate $m_h$ to $m_{A'}$.
Moreover we have shown that inflation generically makes a contribution to the $A'$ density that would overclose the Universe unless $\epsilon$
satisfies a 
{\it lower} bound $\epsilon \gtrsim 10^{-12}$, due to the relation
$m_h/m_{A'}\lesssim 0.01/\epsilon$.
These new bounds disfavor a large region of the previously allowed parameter space, except in the dark photon mass range $10^{-5}$ eV $\lesssim m_{A^{\prime}} \lesssim 1$ eV, the so-called ``direct detection triangle", which is blind to haloscopes and direct detection experiments sensitive to dark photon absorption.

We further derived new constraints on dark photon-mediated thermal dark matter, in the regime where its mass is less than $m_{A'}$,
using the one-loop estimate for $\epsilon$ and constraints on self-interacting dark matter from the Bullet cluster. These considerations rule out $m_{A'} < 46\,$MeV, and place strong (model-independent) upper and (model-dependent) lower bounds on the $\epsilon$ that can diminish the allowed parameter space significantly.

\bigskip
{\bf Acknowledgements.}  
We thank Gonzalo Alonso-\'Alvarez, Andrea Caputo, Miguel Escudero, Andrew Long, Katie Mack, and Michele Redi for helpful discussions.
We are grateful to the CERN Theory Department for its hospitality and financial support during our visits. JC is funded by the Natural Sciences and Engineering Research Council (NSERC) of Canada. GH is supported by the the U.S.\ Department of Energy under award number DE-SC0020250 and DE-SC002026.

\bibliography{ref}
\bibliographystyle{utphys}

\end{document}